\documentclass[%
 reprint,
 superscriptaddress,
 amsmath,amssymb,
 aps,
 prb,
]{revtex4-2}

\usepackage[final]{graphicx}
\usepackage{dcolumn}
\usepackage{bm}
\usepackage{xcolor}
\usepackage{color,soul}
\usepackage{algorithm}
\usepackage{algpseudocode}
\usepackage{url} 
\usepackage{makecell}
\usepackage{multirow}
\usepackage{booktabs}
\usepackage{siunitx}
\DeclareSIUnit\atom{atom}

\usepackage{subfig}
\usepackage{changes}
\usepackage[normalem]{ulem}
\usepackage{hyperref}
\hypersetup{colorlinks = true, allcolors = .}
\usepackage{minted}
\usepackage{orcidlink}

\setcitestyle{super,open={},close={}}
\renewcommand{\textcite}[1]{[\citenum{#1}]}

\begin{document}


\title{Hydride formation and phase separation in palladium nanoparticles from a transferable atomic cluster expansion potential}


\author{Minaam Qamar \orcidlink{}} 
\email[]{minaam.qamar@ruhr-uni-bochum.de}
\affiliation{ICAMS, Ruhr-Universit\"at Bochum, 44801 Bochum, Germany}

\author{Apinya Ngoipala \orcidlink{}}
\email[]{Apinya.Ngoipala@ul.ie}
\affiliation{Materials and Catalysis Modelling Group, Bernal Institute, University of Limerick, Limerick V94 T9PX, Ireland}

\author{Matous Mrovec \orcidlink{}} 
\email[]{matous.mrovec@ruhr-uni-bochum.de}
\affiliation{ICAMS, Ruhr-Universit\"at Bochum, 44801 Bochum, Germany}

\author{Matthias Vandichel \orcidlink{}}
\email[]{Matthias.Vandichel@ul.ie}
\affiliation{Materials and Catalysis Modelling Group, Bernal Institute, University of Limerick, Limerick V94 T9PX, Ireland}

\author{Ralf Drautz \orcidlink{}} 
\affiliation{ICAMS, Ruhr-Universit\"at Bochum, 44801 Bochum, Germany}

\date{\today}

\begin{abstract}

The palladium–hydrogen system is a prototype for hydrogen–metal interactions and underpins technologies such as hydrogen storage, catalysis and purification. Yet its nanoscale behaviour — where surface and interface energetics, elastic coherency strain and size-dependent thermodynamics govern phase separation — has eluded accurate atomistic simulation. Empirical potentials misrepresent the energetics of interstitial hydrogen, while existing machine-learning models are restricted to bulk phases at low-hydrogen environments. Here we introduce an atomic cluster expansion (ACE) for Pd–H that reproduces formation energies, phonon spectra, elastic constants, hydrogen migration barriers and surface adsorption with near-DFT accuracy, benchmarked directly against neutron-scattering, high-pressure and lattice-expansion experiments. Its near-linear scaling and CPU efficiency make molecular dynamics of PdH$_x$ nanoparticles exceeding 28,000 atoms ($\sim$12 nm in diameter) tractable over nanosecond timescales. These simulations resolve, at the atomic scale, the kinetic separation of $\alpha$- and $\beta$-PdH$_x$ into a core–shell architecture, reproduce the experimentally observed size dependence of the lattice parameter, and uncover a pronounced hydrogen-induced lowering of the nanoparticle melting temperature. The potential brings experimentally relevant scales of metal-hydride dynamics within quantitative reach.

\end{abstract}

\maketitle

\section{Introduction}\label{sec:intro}

Understanding the interaction of hydrogen with transition metals is of fundamental and practical interest for both hydrogen evolution and storage, which are the main pillars of the hydrogen-based economy.~\cite{crabtree2004hydrogen,abe2019hydrogen} Palladium (Pd) is widely used to study metal-hydrogen interactions due to its unique capacity to absorb large concentration of hydrogen under practically accessible temperatures and pressures.~\cite{manchester1994h,adams2011role,dekura2019hydrogen} This enables the application of Pd in several hydrogen-related technologies such as hydrogen storage,~\cite{yamauchi2008nanosize,li2014shape} hydrogenation catalysis,~\cite{lopez2006palladium,mccue2015recent,https://doi.org/10.1002/anie.201610552} electrocatalysis,~\cite{benck2019producing,sun2021bifunctional,https://doi.org/10.1002/anie.201914335} and hydrogen purification.~\cite{checchetto2004palladium,rahimpour2017palladium}

Hydrogen absorption in Pd is a reversible process that proceeds via the dissociation of H$_{2}$ molecules on Pd surfaces~\cite{jewell2006review,kaur2019review,somo2020comprehensive} followed by the diffusion of adsorbed H atoms into subsurface layers. In the face-centered cubic (fcc) lattice of bulk Pd, H atoms  predominantly occupy the octahedral interstitial sites. With increasing H content or under specific structural constraints, both octahedral and tetrahedral sites can be populated. When concentration reaches a critical threshold, Pd hydrides (PdH$_{x}$) are formed.~\cite{flanagan1991palladium,worsham1957neutron,cser2004neutron,tew2009particle} Depending on the temperature and H$_2$ partial pressure, two distinct PdH$_x$ phases coexist. 
At low H chemical potentials, a dilute solid solution of H in fcc Pd forms, termed the $\alpha$-phase (approximately $x < 0.03$). As the H content increases, the system transitions into a concentrated $\beta$-hydride phase (with $x > 0.60$).  A miscibility gap exists in the intermediate region ($0.03 < x < 0.60$), where the $\alpha$ and $\beta$ phases coexist. This transition is characterized by a pressure plateau in the pressure-composition isotherms.~\cite{jamieson1976beta, dekura2019hydrogen, goods1992mechanical, griessen2016thermodynamics}

Extensive research effort has been devoted to elucidating the influence of Pd-H composition, temperature, H chemical potential, microstructure and synthesis methods on the H absorption mechanism, hydride stability, and the physical and chemical properties of Pd hydrides. These studies serve to optimize the design principles of Pd materials for hydrogen-related applications.~\cite{benck2019producing, aleksandrov2014absorbed, johansson2010hydrogen, johnson2019facets, suleiman2005hydrogen, landers2021dynamics, https://doi.org/10.1002/anie.202006562, jewell2006review, lischka2003hydrogen} The interaction of H with Pd surfaces has been investigated using various experimental techniques, including low-energy electron diffraction (LEED)~\cite{PhysRevB.40.891,BEHM1980320,10.1063/1.453495}, electron energy-loss spectroscopy (EELS)~\cite{NETZER198326}, thermal desorption spectroscopy (TDS),~\cite{CONRAD1974435} and electrochemical scanning tunneling microscopy (EC-STM).~\cite{https://doi.org/10.1002/smll.202202410}  Furthermore, numerous first-principles calculations based on density functional theory (DFT) have provided insights into the properties of bulk Pd hydrides and the interactions of H with Pd surfaces.~\cite{https://doi.org/10.1002/smll.202202410,hong2007adsorption,abraham2012ab,gross2011coverage,gronbeck2011effect,dong1998hydrogen,ngoipala2024_Htrap,lipin2024computational} However, despite significant advances in high-throughput DFT calculations, inherent limitations remain regarding accessible simulation length and time scales, particularly when investigating kinetics at finite temperatures or systems containing extended defects. 

To address these limitations, various classical interatomic potentials have been developed, such as an embedded atom method (EAM) potential developed by Zhou \textit{et al.}~\cite{zhou_zimmerman_wong_hoyt_2008} or a reactive force field (ReaxFF) potential developed by Senftle \textit{et al.}~\cite{senftle2014reaxff} 
While these classical potentials have been successfully employed in large-scale atomistic simulations of various PdH phases,~\cite{zhou_zimmerman_wong_hoyt_2008, zhou2018molecular, zhou2018temperature, valencia2016hydrogen, crespo2012hydrogen, sun2019atomistic, schwarz2020coherent} they remain inherently limited in their quantitative accuracy due to simplified functional forms upon which they are built. For instance, the EAM potential significantly overestimates the solution energy of an H atom in the tetrahedral sites of fcc Pd (see section~\ref{sec:pd_h_interaction}) and it fails to describe the formation of superabundant vacancy (SAV) phases in the Pd-H system.\cite{kimizuka2022artificial}

During the past decade, machine-learning interatomic potentials (MLIPs) have emerged as highly effective tools for accurately representing the high-dimensional potential-energy surfaces (PES) of diverse material systems. Unlike traditional empirical potentials, MLIPs utilize highly flexible functional forms capable of describing atomic interactions of arbitrary complexity. Consequently, MLIPs serve as efficient surrogate models that, once trained on reliable data generated via DFT, deliver near-DFT accuracy at an orders-of-magnitude reduced computational cost.~\cite{deringer2019machine,behler2016perspective}

Specifically for the Pd-H system, Kimizuka \textit{et al.}~\cite{kimizuka2022artificial} recently developed an artificial neural network (ANN) potential to investigate H diffusion in Pd. Extensive path-integral simulations using this ANN potential revealed subtle differences in temperature-dependent diffusivities of different H isotopes in fcc Pd. However, because this model was trained primarily on bulk configurations with low H content, it is not applicable to studies of defects, such as PdH$_x$ surfaces or Pd/PdH$_x$ interfaces. 

To study the behavior of the Pd-H system in arbitrary configurations and over a broad range of conditions (e.g., temperature, H chemical potential, stress), we have developed a general-purpose atomic cluster expansion (ACE) parametrization.~\cite{Drautz2019_ACE} ACE has been established as one of the most efficient and accurate data-driven interatomic potentials available.\cite{Lysogorskiy2021_npj}. Subsequent developments by Bochkarev \textit{et al.}~\cite{bochkarev2022_prm_pacemaker} and Lysogorskiy \textit{et al.}~\cite{Lysogorskiy_2023_ALace} have significantly streamlined the process of parameterizing and validating ACE potentials. These advancements in potential training and data generation tools~\cite{pyiron_paper} have enabled numerous successful applications across a wide array of materials with diverse bonding characteristics. ACE potentials have seen widespread adoption across various material classes, including metals,~\cite{IbrahimPRM23, NAMAKIAN2023111971, Liang2023,Pan2024} covalent solids,~\cite{qamar2023atomic, erhard2023ACE, Leimeroth2024} liquids and amorphous systems,~\cite{ThomasduToit2024, ATTARIAN2025113409, ATTARIAN2024124521} and even organic materials,~\cite{D4CP01980F}. This underscores their versatility and robustness in extending DFT-level accuracy to large-scale nanoscale modeling.

Standard DFT functionals, such as GGA-PBE~\cite{PBEXC}, are known to exhibit systematic discrepancies for some properties. Specifically for Pd, PBE tends to overestimate the lattice constant and underestimate the bulk modulus due to characteristic underbinding.~\cite{Ilawe2015}
To mitigate these discrepancies, semi-empirical dispersion corrections, such as the D2 or D3 methods by Grimme \textit{et al.}~\cite{Grimme2016}, are frequently employed. While these van der Waals (vdW) corrections often yield better agreement with experiment for equilibrium bulk properties, they may lack transferability across the broad range of configurations encountered in the Pd–H system. For instance, Ilawe et al.~\cite{Ilawe2015} reported that the D2 correction predicts the maximum ideal tensile strength in Pd to occur at strain values nearly double those predicted by other functionals and experimental data. This is fundamentally linked to the limited transferability of the D2 functional, particularly under large stress or structural deformation. The addition of vdW corrections based on the D3 method provides significantly better estimates for these properties because it utilizes geometry-dependent dispersion coefficients.

For the Pd-H system, we developed two ACE potentials using different DFT training datasets. First, we parameterized a more efficient ACE$_{\text{D3}}$ potential trained on data generated with the PBE+D3 correction. This ACE-D3 potential was recently applied successfully to investigate the reconstruction of Pd surfaces under lateral compression induced by hydride formation, as well as PdH$_{x}$ nanoparticles.~\cite{Ngoipala_advMat2025, atlan2025probing,Viola2026} However, even the D3 correction exhibits certain limitations that may impact the predictions of Pd properties far from equilibrium.~\cite{Ilawe2015} Therefore in our second model, we did not include dispersion corrections.
This subsequent model, in the following referred to simply as ACE, was trained on a more extensive dataset and utilized a larger number of basis functions to achieve higher accuracy and transferability.

Recently, foundation models based on equivariant message-passing neural network architectures, such as GRACE, have emerged as general-purpose interatomic potentials capable of describing a wide range of elements and chemical environments within a unified framework.~\cite{bochkarev2024, Lysogorskiy2026} These approaches demonstrate strong accuracy and transferability across diverse systems, making them attractive for broad applications. However, this generality is often accompanied by increased computational cost, as such models typically rely on GPU-accelerated implementations and can be less efficient for large-scale or long-timescale simulations. In contrast, ACE potentials provide a computationally efficient alternative, exhibiting favorable scaling on multi-core CPU architectures and near-linear scaling with system size.~\cite{drautz2019atomic, dusson2022atomic} This makes ACE particularly suitable for extensive molecular dynamics simulations of Pd--H systems while maintaining high accuracy when trained on system-specific datasets.

The manuscript is organized as follows. In section~\ref{sec:results}, we validate the ACE potential against reference DFT for key properties. In section~\ref{sec:applications}, we apply the potentials to study phase-separation in PdH$_x$ nanoparticles and hydrogen-induced surface roughening of Pd/PdH$_x$ surfaces. In section~\ref{sec:conclusions}, we provide an outlook on the current work, along with concluding remarks, and the following section~\ref{sec:moethods} is dedicated to explaining the computational tools used in the study.

\section{Results} \label{sec:results}

The results are presented in two sections. The first section~\ref{sec:eq_props} focuses on near-equilibrium properties of bulk Pd and its hydride phases. These investigations include cohesive and formation energies, basic point defects, and the vibrational and elastic properties. In the second section~\ref{sec:pd_h_interaction}, we examine the interactions of H with bulk fcc Pd and its microstructural features, including diffusion and trapping at point defects and surfaces. The reported results primarily correspond to predictions made by the PBE-based ACE potential.  For comparison, results from the ACE$_\text{D3}$ potential are labeled explicitly where reported.

\subsection{Equilibrium properties} \label{sec:eq_props}

\begin{figure*}[]
    \centering
    {\includegraphics[width=0.9\textwidth]{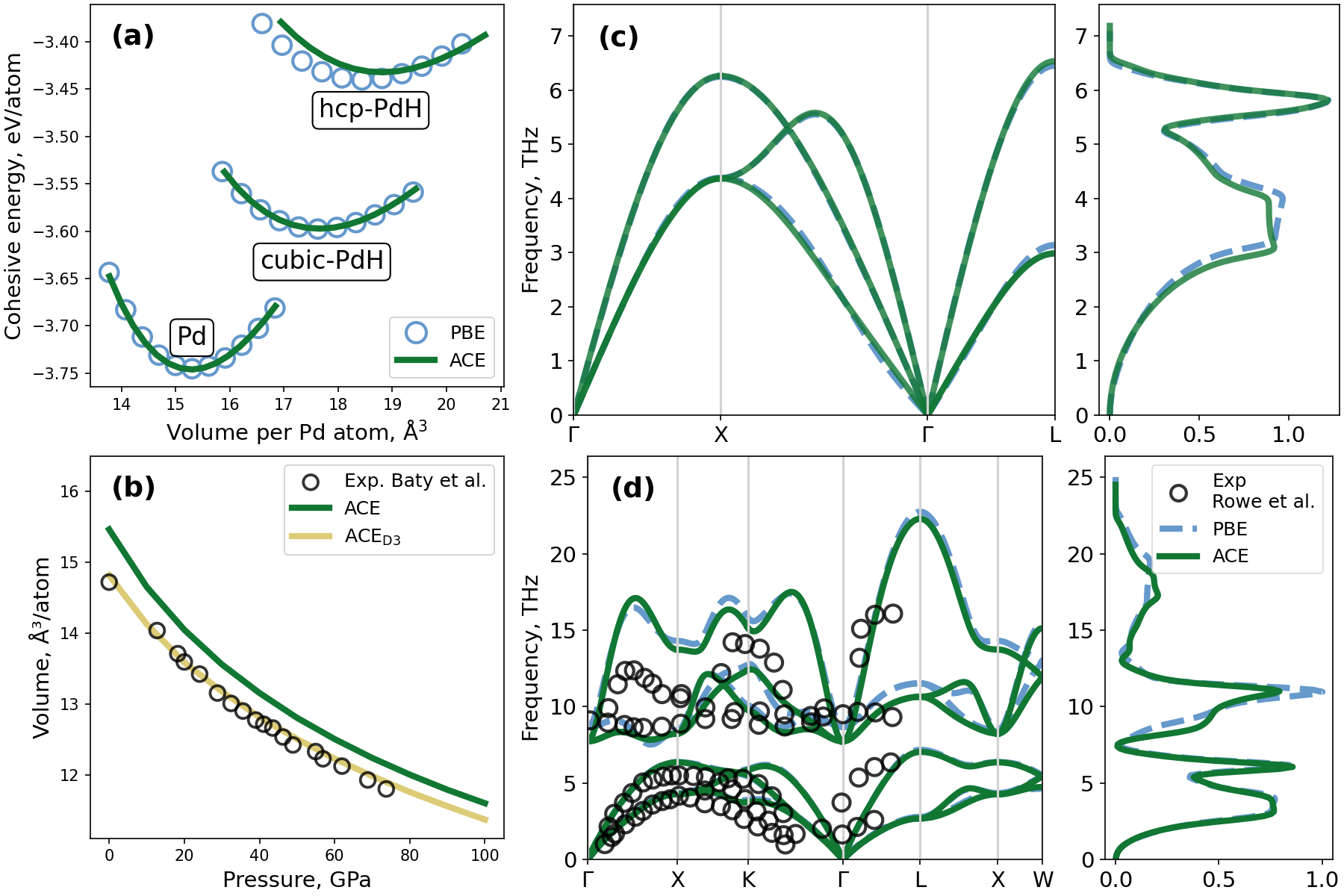}}   
    \caption{
    \textbf{Validation of near-equilibrium properties} (a) Energy-volume curves for fcc Pd and both cubic and hexagonal PdH phases, validated with PBE predictions, (b) atomic volume as a function of external pressure for fcc-Pd as obtained from direct MD simulations, validated with results from isotropic compression experiments reported in Ref.~\cite{baty_et_al2024}, and phonon band structures and associated phonon density of states plotted for (c) equilibrium Pd-fcc, (d)  stoichiometric cubic PdH under 15.2 GPa pressure. Dotted lines represent DFT reference from this work, and points mark experimental data taken from ref.~\textcite{Rowe_PdH_phonons_exp} 
    }%
    \label{fig:basic_validation}%
\end{figure*}

The basic bulk properties of Pd and its hydride phases are presented in Table~\ref{tab:main_props}. The ACE potential predicts the equilibrium lattice constant for fcc Pd of 3.94 \AA, which is in perfect agreement with the reference PBE value. Figure~\ref{fig:basic_validation}(a) shows the near-equilibrium energy-volume curves for fcc Pd, cubic PdH, and hexagonal PdH phases. The results for fcc Pd and cubic PdH agree perfectly with reference DFT calculations. Furthermore, the hexagonal PdH is also modeled accurately, with ACE exhibiting only minor deviations under compression.

The equilibrium volumes of cubic and hexagonal PdH phases are predicted to be 15.4 \% and 21.8 \% larger than that of fcc Pd, respectively. The predicted volume expansion for cubic PdH is consistent with previous DFT calculations by Kimizuka et al.~\cite{kimizuka2022artificial}, who reported a 14.3 \% increase relative to fcc Pd. These results underscore the significant lattice mismatch that occurs during hydride formation. Specifically, when a cubic PdH nucleus is constrained to the lattice constant of the underlying Pd substrate, for example, during surface nucleation, the hydride phase is subjected to substantial compressive stresses.

\begin{table}[]
\begin{tabular}{lcclc}
\hline
\multicolumn{1}{l|}{}                    & PBE    & ACE    & PBE$_{\text{D3}}$ & ACE$_{\text{D3}}$ \\ \hline
\textbf{Pd}                              &        &        &        &                   \\ \hline
\multicolumn{1}{l|}{$a_0$}               & 3.94   & 3.94   & 3.88   & 3.88              \\
\multicolumn{1}{l|}{$E_{coh}^{Pd}$}      & -3.74  & -3.74  & -4.39  & -4.39             \\
\multicolumn{1}{l|}{$C_{11}$}            & 205    & 203    & 228    & 262               \\
\multicolumn{1}{l|}{$C_{12}$}            & 148    & 150    & 173    & 163               \\
\multicolumn{1}{l|}{$C_{44}$}            & 63     & 64     & 80     & 91                \\
\multicolumn{1}{l|}{$E^f_{vac}$}         & 1.14   & 1.14   & 1.56   & 1.79              \\
\multicolumn{1}{l|}{$\gamma^{111}$}      & 1.32   & 1.36   & 2.16   & 2.29              \\
\multicolumn{1}{l|}{$\gamma^{100}$}      & 1.50   & 1.54   & 2.24   & 2.58              \\
\multicolumn{1}{l|}{$\gamma^{110}$}      & 1.61   & 1.60   & 2.4    & 2.64              \\
\multicolumn{1}{l|}{$E^{{sol}}_{octa}$}  & -0.082 & -0.087 & -0.184 & -0.190            \\
\multicolumn{1}{l|}{$E^{{sol}}_{tetra}$} & -0.046 & -0.046 & -0.121 & -0.134            \\ \hline
\textbf{H$_2$ dimer}                     &        &        &        &                   \\ \hline
\multicolumn{1}{l|}{$d_{H_2}$}           & 0.75   & 0.75   &        & 0.75              \\
\multicolumn{1}{l|}{$E_{coh}^{H}$}       & -3.32  & -3.33  &        & -3.32             \\ \hline 
\end{tabular}
\caption{Basic bulk properties of Pd fcc and H$_2$ dimer as predicted by the ACE and ACE$_{\text{D3}}$ potentials and our PBE DFT calculations. The lattice parameter ($a_0$) and the interatomic distance ($d_{H_2}$) are given in \AA{}; cohesive, solution and migration energies are given in eV/atom; surface energies ($\gamma^{hkl}$) are given in $J/m^2$; elastic constants ($C_{ij}$) are in GPa.}\label{tab:main_props}
\end{table}

In a combined experimental and theoretical study, Baty et al.~\cite{baty_et_al2024} studied the response of pure Pd under pressure, evaluating the room temperature equation of state (EOS) for fcc Pd under isotropic compression up to 80 GPa using a diamond anvil. We replicated this experiment by performing direct molecular dynamics (MD) simulations of an 8 $\times$ 8 $\times$ 8 fcc Pd supercell within the $NPT$ ensemble. The predictions from both ACE potentials, along with the experimental data, are presented in Fig.~\ref{fig:basic_validation}(b). As discussed above,~\cite{Ilawe2015} the PBE-based ACE potential exhibits an expected systematic volume shift but accurately reproduces the experimental pressure-volume trend. In contrast, the ACE$_{\text{D3}}$ potential achieves excellent quantitative consistency with experimental observations across the entire pressure range. However, it should be noted that this close agreement might be partly fortuitous.

The phonon band structure and associated density of states for fcc Pd are shown in Fig.~\ref{fig:basic_validation}(c). The ACE potential exhibits excellent agreement with reference DFT calculations. Fully stoichiometric cubic PdH is predicted to be dynamically unstable at zero pressure by the PBE functional. However, it becomes stable under compression.~\cite{ISAEVA_abrikosov2011} Figure~\ref{fig:basic_validation}(d) presents the phonon band structure and associated density of states for stoichiometric cubic PdH at a compression of 15.2 GPa. Experimental data obtained via coherent neutron inelastic scattering of PdD$_{0.63}$ crystals~\cite{Rowe_PdH_phonons_exp} are plotted alongside ACE and reference DFT. The experimental data indicate that the phonon dispersion of PdH exhibits a distinctive energy gap between the optical and acoustic branches at the high-symmetry point $X$, a feature well reproduced by the ACE potential. Although ACE overestimates the optical branch frequencies, this discrepancy is attributed to the use of  deuterium in the experiments. Because deuterium is heavier than hydrogen, the experimentally observed phonon frequencies for PdD$_{x}$ are lower. In hydrides containing hydrogen (protium), a shift toward higher phonon frequencies is expected due to the lower mass of the isotope,~\cite{Rowe_PdH_phonons_exp} which aligns more closely with the ACE predictions.

\subsection{Pd interaction with hydrogen} \label{sec:pd_h_interaction}

\begin{figure*}[]
    \centering
    {\includegraphics[width=0.9\textwidth]{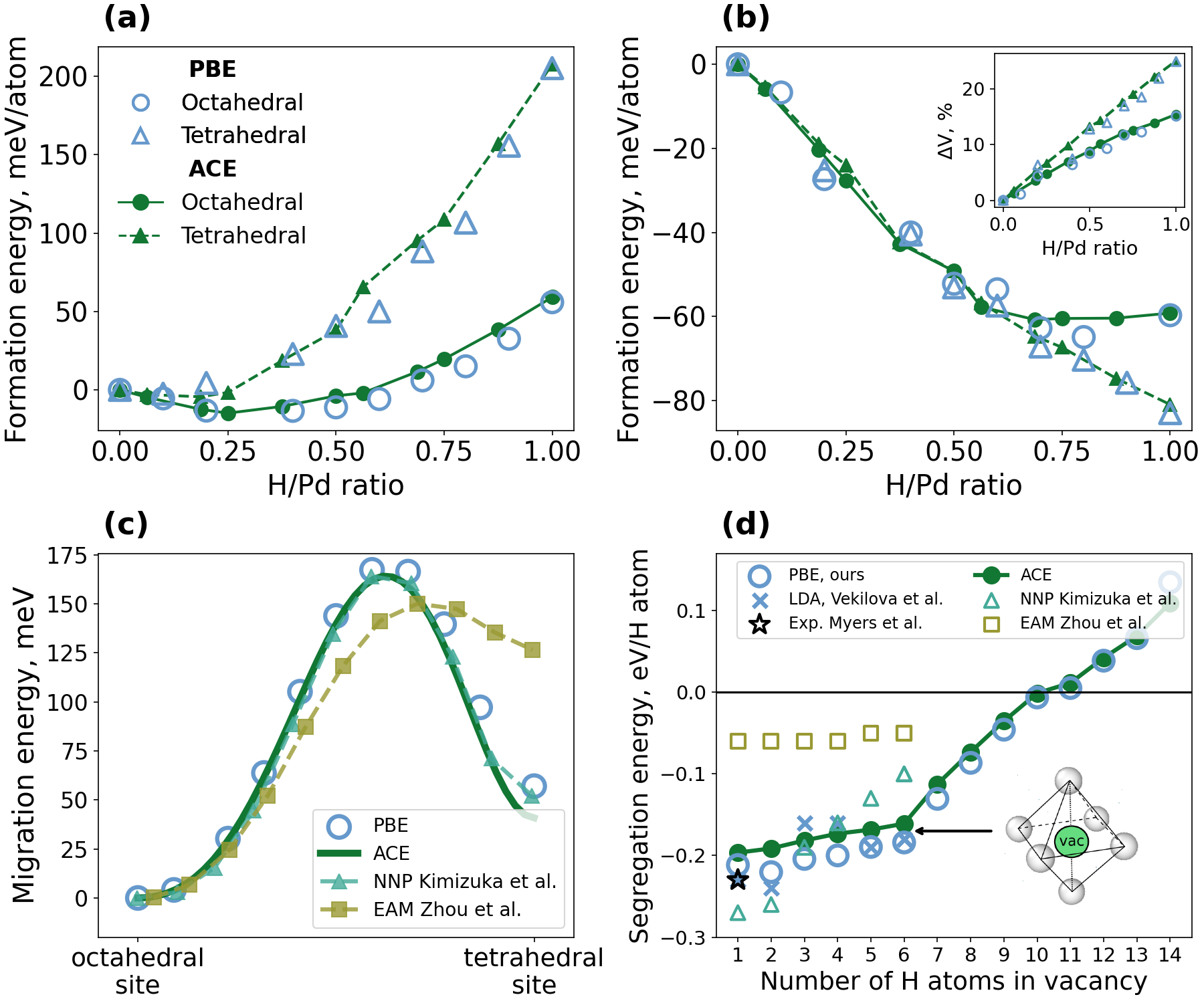}}   
    \caption{
    \textbf{Interaction of H atoms with bulk Pd.} Change in the formation energy of PdH$_{x}$ as a function of H/Pd ratio for (a) fixed volume and (b) isometrically relaxed volumes. The dotted and the solid lines represent structures with H atoms occupying tetrahedral and octahedral sites, respectively. The inset in (b) shows the percentage change in volume of the unit cell. (c) The migration path of a single H atom from an octahedral site to a neighboring tetrahedral site in fcc Pd; and (d) segregation energies of multiple H atoms in a single vacancy in fcc Pd. Experimental data are derived from ion-beam experiments~\cite{Myers_H_vac_EXP}; DFT segregation energies were calculated using the LDA functional~\cite{Vekilova_PdH_vac}. 
    }%
    \label{fig:pdHx_interaction}%
\end{figure*}

The formation of PdH occurs as H atoms start occupying the interstitial sites in fcc Pd. By gradually filling these sites, the entire composition range of the PdH$_{x}$ phases can be simulated. To achieve this, we constructed a 16-atoms fcc-Pd supercell by duplicating the conventional fcc unit cell in a 2 $\times$ 2 $\times$ 1 configuration. Hydrogen atoms were then incrementally added to the interstitial sites until the fully stoichiometric PdH phase was reached. In fcc Pd, hydrogen atoms can occupy either the octahedral (O) or tetrahedral (T) interstitial sites, with the former being the favorable sites at standard conditions and lower H concentrations. At higher H concentrations, the relative stability of the T sites can increase, depending on whether lattice expansion is permitted.~\cite{Kimizuka_2018PRB, akiba2016nanometer} 

To investigate this further, we computed the formation energies of PdH$_{x}$ crystals as a function of H concentration, populating exclusively either the T sites or the O sites. Figure~\ref{fig:pdHx_interaction}(a) illustrates the change in formation energy calculated at the fixed equilibrium volume of pure fcc Pd (only the atomic positions relaxed), while Fig.~\ref{fig:pdHx_interaction}(b) presents the results obtained with full isometric volume relaxation. In all the cases, the ACE predictions demonstrate excellent agreement with the reference PBE calculations. Under fixed volume conditions, the formation energies of PdH$_x$ with H occupying the T sites rise more rapidly than those occupying the O sites. This is consistent with previous observations by Zhou et al.~\cite{zhou_zimmerman_wong_hoyt_2008}  We note that at high H ratios ($x>0.55$ for O sites and $x>0.3$ for T sites), the formation energies become positive, indicating that the hydride phase is under compressive strain. When the lattice volume is allowed to relax, the O sites exhibit lower formation energies only at lower H concentrations ($x < 0.55$). At higher H concentration, the formation energy of hydrogen in the T sites decreases rapidly, eventually making these sites energetically more favorable than the O sites. The inset in Fig.~\ref{fig:pdHx_interaction}(b) shows the percentage change in volume. Notably, for H atoms in the T sites, the lattice is expected to expand as much as 25 \%.

Nudged-elastic band (NEB) calculations were performed to calculate the minimum energy path for migration of an H atom from an O site to a neighboring T site. The migration energy is plotted in Fig.~\ref{fig:pdHx_interaction}(c). The ACE potential predicts a migration barrier of 163 meV, in perfect agreement with reference DFT calculations. The difference between the solution energies of H in the O and T sites is calculated to be 48 meV, closely matching the 53 meV predicted by reference PBE. While the NNP potential developed by Kimizuka et al.~\cite{kimizuka2022artificial} reproduces the entire path perfectly, the classical EAM potential developed by Zhou et al.~\cite{zhou_zimmerman_wong_hoyt_2008} highly overestimates the solution energy of H atom in the T site.

The formation energy of a monovacancy in a 5 $\times$ 5 $\times$ 5 fcc Pd supercell is predicted to be 1.141 eV. This value is in close agreement with our reference DFT calculations (1.145 eV) and the DFT value (1.15 eV) computed by Kimizuka et al.~\cite{kimizuka2022artificial} The small atomic radius of hydrogen allows multiple H atoms to occupy trapping sites around a single Pd vacancy. Earlier DFT calculations performed by Vekilova et al.~\cite{Vekilova_PdH_vac} using the LDA functional calculated the segregation of up to six H atoms in the octahedral sites surrounding a single vacancy. Interestingly, the center of the vacancy site is not favourable and the H atoms tend to occupy the O or T sites neighbouring the vacancy. The six O sites surrounding the monovacancy, forming a square bipyramid [see the inset in Fig.~\ref{fig:pdHx_interaction}(d)], are the most stable trapping sites. As shown in Fig.~\ref{fig:pdHx_interaction}(d), the segregation of up to six H atoms into these sites is energetically favorable, with the segregation energy per H atom increasing gradually from $-0.19$ to $-0.16$ eV/H atom. 

Beyond these six O sites, additional 8 H atoms can be accommodated in the neighboring T sites, resulting in a total of 14 H atoms clustered around a single Pd vacancy. However, populating these T sites causes the segregation energy to rise sharply, becoming positive (unfavourable) for more than 10 H atoms. 

For comparison, Fig.~\ref{fig:pdHx_interaction}(d) includes results from ion-beam experiments by Myers et al.~\cite{Myers_H_vac_EXP} for a single H atom in a vacancy, along with predictions from other interatomic potentials. The segregation energies obtained by ACE agree well with our reference PBE results and the LDA values reported by Vekilova et al.~\cite{Vekilova_PdH_vac} for up to six H atoms. In contrast, the segregation energies predicted by the NNP potential decrease more sharply than DFT reference, while the EAM potential significantly underestimates the segregation energies by more than a factor of two.

\begin{figure*}[]
    \centering
    
    \includegraphics[height=0.9\textheight,keepaspectratio]{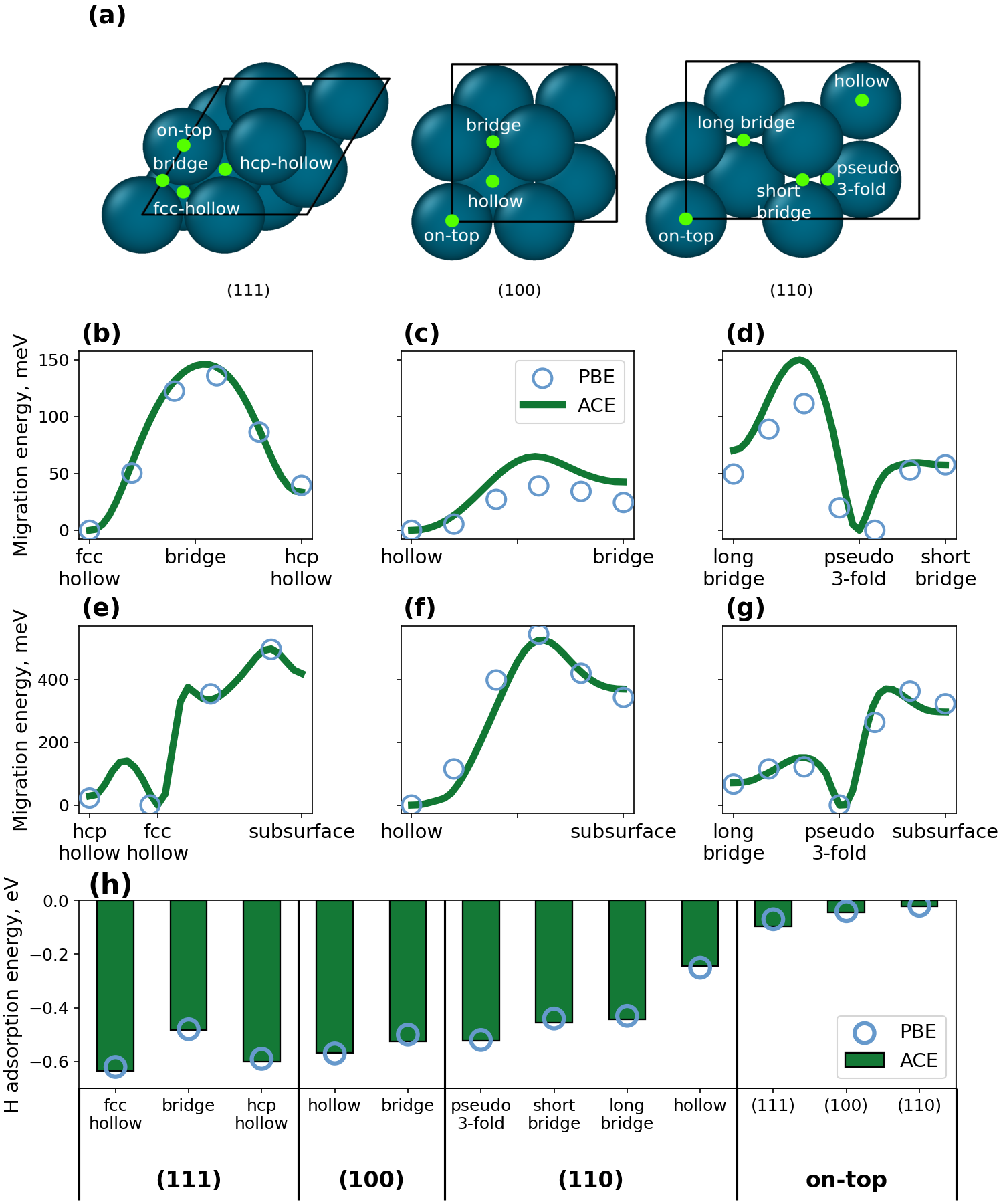} %
    
    \caption{\textbf{Migration and adsorption energies of H atoms on selected Pd surfaces.} (a) Schematic illustration of key adsorption sites for H atoms on low-index Pd surfaces. Panels (b)–(g) show the computed migration energy barriers obtained via NEB calculations: panels (b)–(d) correspond to H migration between surface adsorption sites, while panels (e)–(g) illustrate H migration from surface sites into subsurface octahedral sites for the (111), (100), and (110) surfaces, respectively. In the above context, we use subsurface for the first subsurface layer. Panel (h) summarizes H adsorption energies at the depicted surface sites (at a coverage of 25\%), calculated relative to H in the gas phase.
    }%
    \label{fig:surf_mig_traps}%
\end{figure*}

Each surface orientation provides unique adsorption sites for H atoms, characterized by distinct coordination environments. The primary trapping sites for the (111), (100) and (110) surfaces are illustrated in Fig.~\ref{fig:surf_mig_traps}(a). Due to the relatively strong interaction between Pd and H atoms, the stability of these adsorption sites is generally correlated with the coordination number of H atoms to neighboring Pd atoms. 

On the (111) surface, two types of threefold-coordinated adsorption sites exist: the `fcc-hollow' and `hcp-hollow' sites, named according to their respective stacking sequences. In contrast, the (100) surface features a fourfold-coordinated `hollow' site, while the (110) surface contains a twofold-coordinated `hollow' site. Across all surface orientations, the `bridge' sites exhibit twofold coordination, whereas the `on-top' sites involve a single Pd atom, resulting in single coordination. An additional stable site exists on the (110) facet, when an H atom moves from a hollow site toward the short bridge. The so-called pseudo-threefold site features higher coordination and is consequently more stable than the standard (110) hollow site.

A 4 $\times$ 4 surface supercell is used to construct the (100) facet which has eight exposed Pd atoms. 
This corresponds to four available hollow, bridge and on-top sites. A single H atom adsorbed on one of the sites thus corresponds to H coverage of 25 \%.  Similarly, 2 $\times$ 2 orthorhombic surface supercells are used for the (111) and (110) facets, each corresponding to 25 \% H coverage with a single H atom occupying one of the surface sites.

We calculated the adsorption energies of H atom on various Pd surface trapping sites to quantify their relative stabilities. The adsorption energies, $E_{ads}$, were calculated relative to the H$_{2}$ molecule according to the following expression:

\begin{equation}
    E_{ads} = E_{surface}^{H_2} - ( E_{surface} + E_{coh}^{H_2} ) ,
\end{equation}

where $E_{surface}^{nH}$ represents the energy of the surface slab with $n$ adsorbed H atoms, $E_{surface}$ is the energy of the pristine surface slab, and $E_{coh}^{H_2}$ is the binding energy of the hydrogen dimer  (cf. Table~\ref{tab:main_props}). A negative value of $E_{ads}$ indicates exothermic adsorption, with more negative values signifying stronger adsorption. 

The adsorption energies predicted by the ACE potential for various Pd surface sites are presented in Fig.~\ref{fig:surf_mig_traps}(h), together with DFT reference data. The surface energies of the three low-index facets, reported in Table~\ref{tab:main_props}, follow the energetic ordering $\gamma^{111} < \gamma^{100} < \gamma^{110}$. This correlates with the observed H adsorption energies, where sites on the (111) surface exhibit the lowest adsorption energies. 

On the (111) surface, the threefold fcc-hollow site is the most stable with the adsorption energy of $-0.634$ eV per H atom, followed closely by the twofold bridge ($-0.633$ eV), hcp-hollow ($-0.62$ eV), and on-top ($-0.09$ eV) sites. For the (100) surface, the fourfold hollow site is the most favourable ($-0.57$ eV), followed by the bridge site ($-0.52$ eV) and on-top ($-0.04$ eV) sites. On  the (110) surface, the standard hollow site is only metastable and a slight perturbation causes the H atom to migrate toward the more stable pseudo-threefold site ($-0.52$ eV). Other sites on this surface facet follow the sequence: shortbridge ($-0.45$ eV), longbridge ($-0.44$ eV), hollow ($-0.24$ eV), and on-top ($-0.02$ eV).

The diffusion of H atoms on and into Pd surfaces is governed by distinct energetic pathways across different facets. We performed further NEB calculations to determine the minimum energy paths for H migration between various surface trapping sites. In general, surface migration occurs with significantly lower transition barrier compared to bulk diffusion. 

The NEB pathways for surface migration across the (111), (100), and (110) Pd surfaces are shown in Figs.~\ref{fig:surf_mig_traps}(b)-(d), respectively. The largest barrier on the (111) surface occurs when an H atom migrates from an fcc-hollow site to an hcp-hollow site, via the two-fold bridge site, incurring a barrier of 149 meV. On the (110) surface, starting from the stable pseudo-threefold site, the H atom can migrate to the short-bridge site ($\Delta$E$_{mig}$ = 56 meV) or the longbridge site ($\Delta$E$_{mig}$ = 147 meV). For the (100) surface, the migration from a hollow site to a bridge site has a small migration barrier of 58 meV. While these values represent the MEP, higher-energy pathways also exist, such as migration via the energetically unfavorable on-top sites.

As surface sites saturate, H atoms begin to diffuse into the subsurface layers of the Pd lattice. The NEB paths for surface-to-subsurface migration for the (111), (100), and (110) facets are shown in Fig.~\ref{fig:surf_mig_traps}(e)-(g), respectively. In each case, we examined the transition from the most stable surface site to a subsurface octahedral site. For the (111) and (100) surfaces, the migration barriers are 0.52 eV and 0.53 eV, respectively. The subsurface migration in the (110) facet proceeds via the pseudo-threefold site with a notably smaller barrier of 0.39 eV.

\begin{table}[]
\begin{tabular}{l|ccc}
\hline
Surface orientation & DFT                 & ACE             & ACE$_{\text{D3}}$      \\ \hline
(111)               & $-0.55$ $\pm$ 0.05$^a$ & $-0.51$ $\pm$ 0.02 & $-0.50$ $\pm$ 0.01 \\
(110)               & $-0.46$ $\pm$ 0.06$^b$ & $-0.35$ $\pm$ 0.06 & $-0.58$ $\pm$ 0.09 \\
(100)               & $-0.46$ $\pm$ 0.03$^c$ & $-0.42$ $\pm$ 0.04 & $-0.58$ $\pm$ 0.07 \\ \hline
\end{tabular} \\

$^a$ Ref.~\citenum{ROUDGAR2003121}, $^b$ Ref.~\citenum{Ledentu1998}, $^c$ Ref.~\citenum{gross2011coverage}

\caption{Mean H$_2$ dissociation and adsorption energies (in eV per H atom) for different Pd surfaces.
}\label{tab:H_adsorption_energies}
\end{table}

\begin{figure}[]
    \centering
    {\includegraphics[width=0.95\columnwidth]{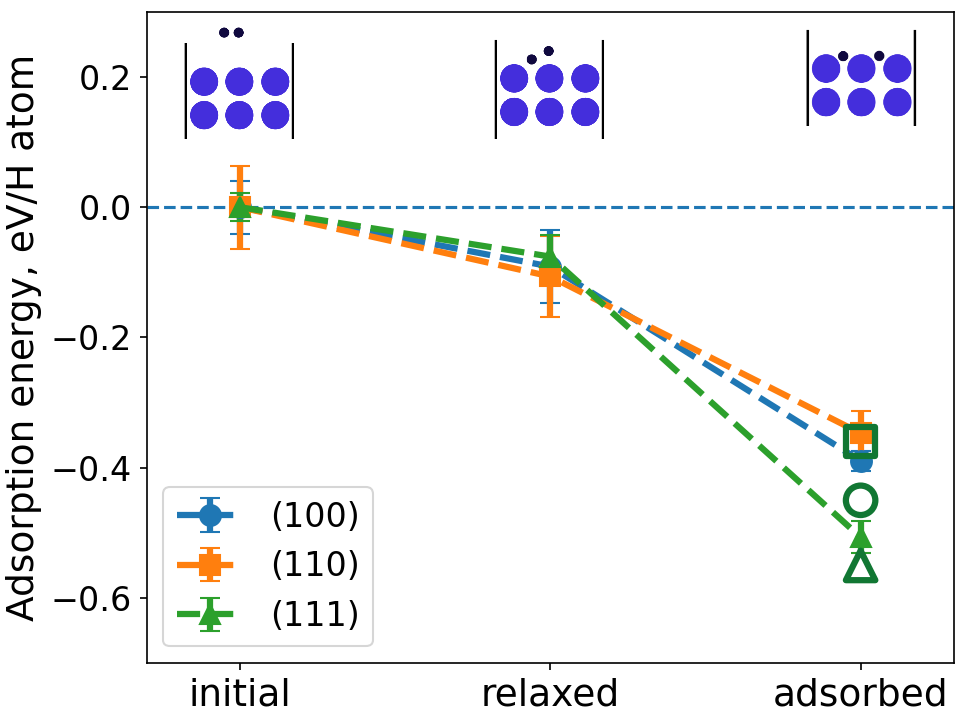}}   
    \caption{
    H adsorption energy on Pd surfaces from initial H$_2$ placement to relaxed and fully adsorbed states. Averaged values over multiple trajectories are shown with error bars reflecting site-dependent variability.
    }%
    \label{fig:h2_on_surf}%
\end{figure}

Upon contact with Pd surfaces, H$_2$ gas molecules dissociate into individual H atoms, which subsequently diffuse and adsorb on the Pd surface. This dissociation process is an important step that precedes the formation of hydride and the ACE potential was trained to reproduce the DFT-derived pathways for H dissociation and absorption. 

To quantify the adsorption energetics, a single H$_2$ molecule was initially placed 2~\AA{} above the Pd surface. After a short $NVT$ MD run (10 ps at 300 K) and subsequent structural relaxation, the H atoms were found to diffuse into either the surface or subsurface layers of the Pd slab. Given the stochastic nature of the dissociation mechanism, the resulting absorption energy varies based on the initial placement of the H$_2$ molecule. Consequently, the process was repeated by placing the molecule above different adsorption sites for all surface orientations. The mean energies for each orientation are presented in Table~\ref{tab:H_adsorption_energies} and Fig.~\ref{fig:h2_on_surf}.

The largest energy gain is observed for H$_2$ dissociation on the Pd(111) surface, yielding an adsorption energy of 0.51 $\pm$ 0.02 eV per H atom. This value is in close agreement with 0.55 $\pm$ 0.05 eV per H atom obtained from PBE-DFT calculations by Roudgar et al.~\cite{ROUDGAR2003121}, and 0.65 eV per H atom reported in earlier PW91-based DFT calculations by Senftle et al.~\cite{senftle2014reaxff}. For the Pd(110) surface, however, ACE slightly underestimates the adsorption energy compared to the corresponding DFT reference values.

\section{Application: Nanoparticles} \label{sec:applications}

The results reported above validate the ability of the ACE potential to reproduce critical properties of the Pd-H system with near-DFT accuracy. In this section, we apply the potential to studies of PdH$_x$ nanoparticles at finite and high temperatures. Such simulations present significant challenges for MLIPs, as they probe highly nonequilibrium regions of PES. Consequently, performing these simulations requires potentials that accurately describe atomic interactions far from equilibrium while maintaining the computational efficiency necessary to simulate large-scale systems over nanosecond timescales.

\subsection{Characterization in PdH$_x$ nanoparticles}


\begin{figure*}[]
    \centering
    {\includegraphics[width=0.87\textwidth]{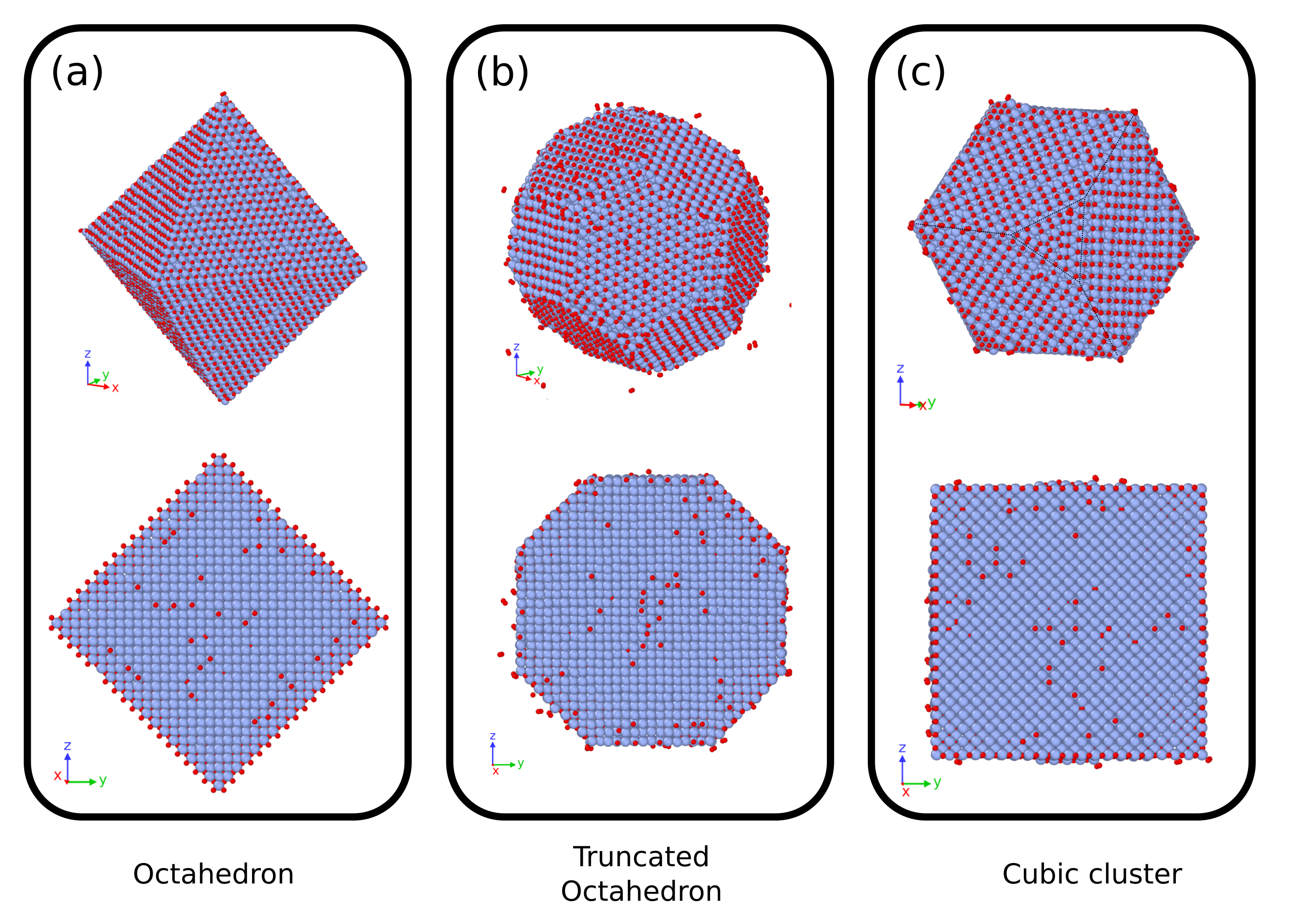}}   
    \caption{
    Snapshots of equilibrated PdH$_{0.3}$ nanoparticles with different geometries. The top row shows 3D perspective views of the nanoparticles, the bottom row their cross-sections.
    }%
    \label{fig:nanoclus_schematic}%
\end{figure*}

The phase separation mechanisms in finite-size PdH$_x$ nanoparticles remain a subject of intense active research, as they are governed by a subtle competition between surface and interface energetics, elastic coherency strain, and size-dependent thermodynamics.~\cite{schwarz2006thermodynamics, baldi2014situ, griessen2016thermodynamics, syrenova2015hydride, ulvestad2017self, ulvestad2017three, sun2019atomistic, schwarz2020coherent} Capturing this interplay demands simulations at realistic length and time scales, which are enabled here by the computational efficiency and linear scaling of the ACE potential, allowing us to simulate nanoparticles consisting of tens of thousands of atoms over nanosecond timescales using modest computational resources.

The largest system modeled in this study was a cubic cluster approximately 12 nm in diameter, containing 28,383 Pd atoms — a size that surpasses the smallest nanoparticles routinely synthesized in experimental laboratories.  For reference, Moumaneix and coworkers~\cite{Moumaneix_ChemElectroChem2023} synthesized Pd nanoparticles with diameters ranging from 3 to 25 nm to study the effect of nanoparticle size on absorbed H concentration, while in recent work Schott et al.~\cite{Schott_Small2024} successfully synthesized Pd nanoparticles as small as 6.4 nm using a novel top-down fabrication approach. 

For the initial investigation of nanoparticles, we modeled several experimentally relevant nanoparticle geometries~\cite{Zhang2001, Saldan2025} across a range of PdH$_{x}$ compositions. Specifically, we considered three distinct structural archetypes: (a) an octahedral cluster with (111) facets containing 9,224 Pd atoms, (b) a truncated octahedral cluster based on the Wulff constructions that accounts for the  surface-energy hierarchy ($\gamma^{111} < \gamma^{100} < \gamma^{110}$) containing 8,439 Pd atoms, and (c) a capped cubic cluster primarily bounded by (100) facets truncated by (111) facets at the corners containing 8,621 Pd atoms. Visual representations of all three cluster geometries are presented in Fig.~\ref{fig:nanoclus_schematic}.

The initial Pd clusters had approximate diameters of 8.8 nm (octahedral), 6.5 nm (truncated octahedral), and 7.8 nm (cubic). For each geometry, we explored three hydrogen concentrations to characterize the system's behavior across the phase diagram: (i) PdH$_{0.03}$, representing the dilute $\alpha$ phase, (ii) PdH$_{0.3}$ corresponding to the $\alpha - \beta$ coexistence region, and (iii) PdH$_{0.6}$ representing the hydride $\beta$ phase. Initially, H atoms were distributed randomly throughout the interstitial sites of each nanoparticle. All structures underwent an initial structural optimization, followed by MD simulations in the $NVT$ ensemble at 300 K for 3 ns. This timescale was chosen to reach equilibrium for each concentration and geometry.

\begin{figure}[]
    \centering
    {\includegraphics[width=0.48\textwidth]{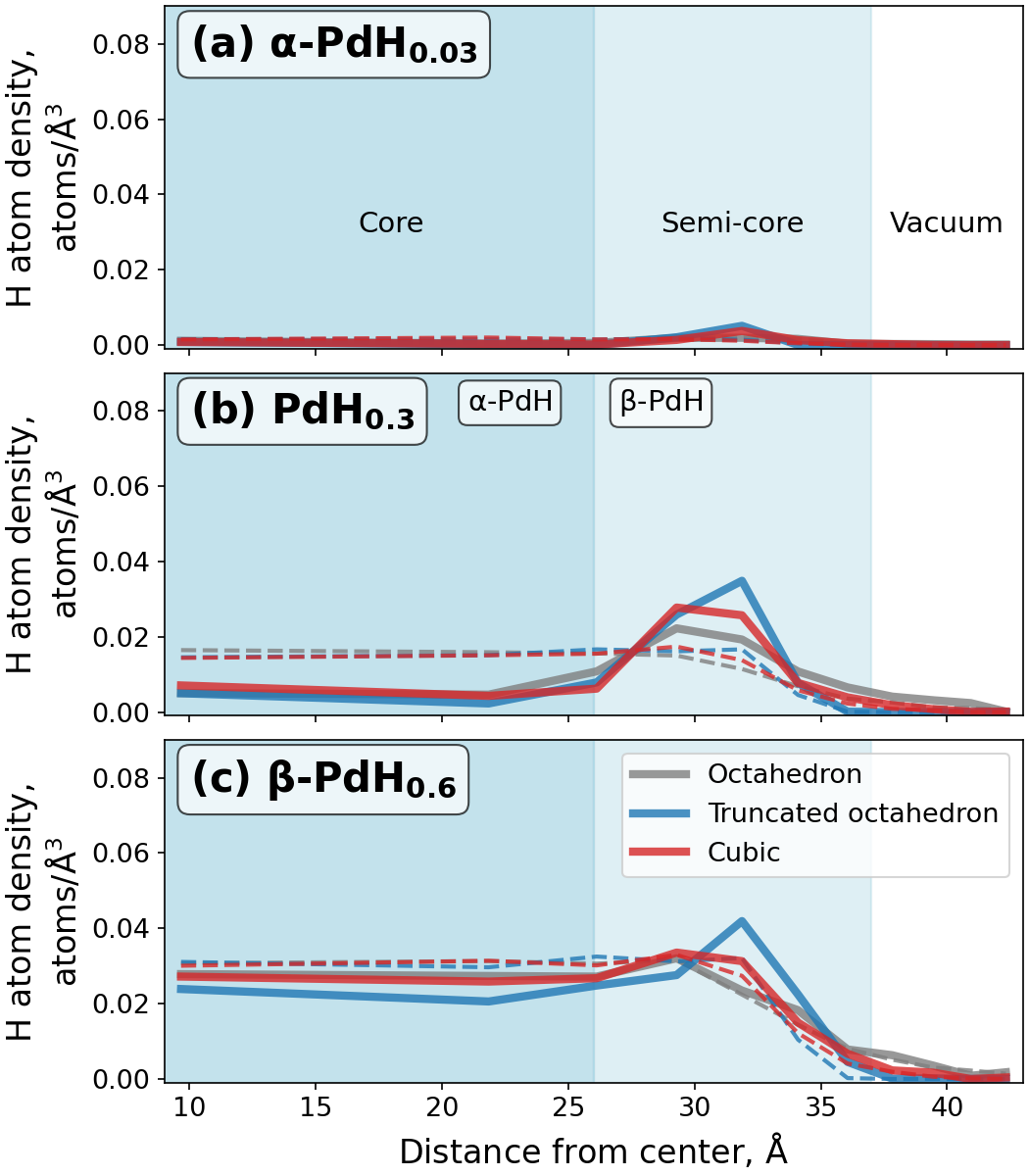}}   
    \caption{
    Distribution of H atoms in the PdH$_x$ nanoclusters of three different shapes and H concentrations. Dashed lines indicate the initial (pre-equilibration) distributions. Shaded regions denote core, semi-core, and vacuum regions.
    }%
    \label{fig:H_distr_}%
\end{figure}

During the MD simulation, H atoms diffuse through the Pd lattice to sample and occupy more energetically favourable positions. The equilibration time of 3 ns was found sufficient to reach a steady state with H atoms fully redistributed within the nanoparticles. The resulting radial density distribution of H and Pd atoms is displayed as a function of distance from the particle center in Fig.~\ref{fig:H_distr_}. For the dilute $\alpha$-PdH$_{0.03}$ phase, H atoms segregate exclusively to the nanoparticle surfaces, leaving the cluster core practically H free. In the mixed PdH$_{0.3}$ phase, equilibration leads to a distinct phase separation: the majority of H atoms migrate to the surface, saturating it into a $\beta$-PdH shell, while a smaller fraction remains within the core, forming a dilute $\alpha$-PdH region. For the fully hydrated $\beta$-PdH$_{0.6}$ phase, the H atoms are uniformly distributed throughout the core, resulting in a coherent hydride crystal. Snapshots of the equilibrated structures for the  PdH$_{0.3}$ phase are shown in Fig.~\ref{fig:nanoclus_schematic}; the top row showing a 3D perspective view of the nanoparticles, the bottom row provides a cross-sectional view of the cores.

\begin{figure}[]
    \centering
    {\includegraphics[width=0.97\columnwidth]{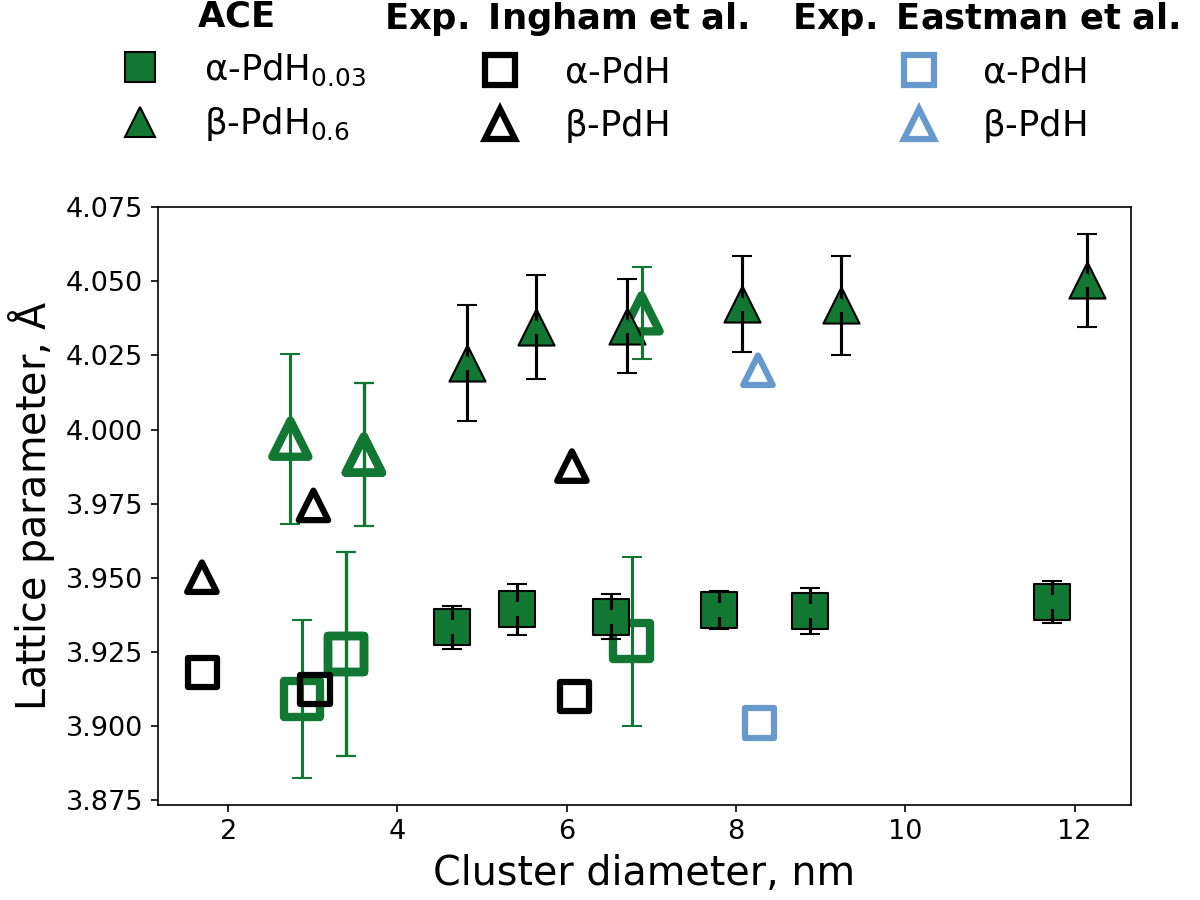}}   
    \caption{Approximate lattice parameter of PdH$_x$ nanoparticles as a function of cluster diameter for two hydrogen concentrations: $\alpha$-PdH$_{0.03}$ (squares) and $\beta$-PdH$_{0.6}$ (triangles). Experimental reference data are taken from Ingham \textit{et al.}~\cite{Ingham2008} and Eastman \textit{et al.}~\cite{Eastman1993}. Open green symbols denote clusters generated with ACE using a melt--quench protocol.}
    \label{fig:cluster_size}%
\end{figure}

A distinguishing feature of the $\alpha$ and $\beta$ phases is their significant difference in their respective lattice parameters. To investigate this, we modeled a series of PdH$_{x}$ nanoparticles in both the $\alpha$ and $\beta$ phases across a range of diameters. To determine the effective lattice parameter for each nanoparticle, the structures were first equilibrated at 300 K until the H atoms were fully redistributed, followed by structural relaxation. All nearest-neighbor Pd-Pd distances were computed, and the resulting average lattice constants are shown in Fig.~\ref{fig:cluster_size} as a function of nanoparticle size, alongside experimental measurements by Ingham et al.\cite{Ingham2008} and Eastman et al.~\cite{Eastman1993}

Since the simulations were performed using the PBE-based ACE potential, a systematic upward shift of the computed lattice parameters is observed relative to the experimental values, consistent with the characteristic underbinding of the PBE functional. Furthermore, the computed lattice parameters exhibit a standard deviation that reflects the differential expansion between the near-surface region, which can expand more freely, and the core region, whose expansion is constrained. This effect is particularly pronounced when comparing error bars between the $\beta$ and $\alpha$ particles. Due to the significantly greater lattice expansion of the shell region in the $\beta$ phase (cf. Section~\ref{sec:pd_h_interaction}, Fig.~\ref{fig:pdHx_interaction}(b)), the lattice constants for nanoparticles in the $\beta$ phase have larger errors, compared to those in the $\alpha$ phase.
A subset of the nanoclusters was generated using melt–quench simulations (see section~\ref{sec:superheated_cluster}). Their lattice constants are indicated by hollow symbols in Fig.~\ref{fig:cluster_size}. Rapid quenching introduces significant structural disorder, leading to a larger deviation in the average lattice constants for both the $\alpha$ and $\beta$ phases.

\subsection{Superheated nanoparticles} \label{sec:superheated_cluster}

\begin{figure*}[]
    \centering
    {\includegraphics[width=\textwidth]{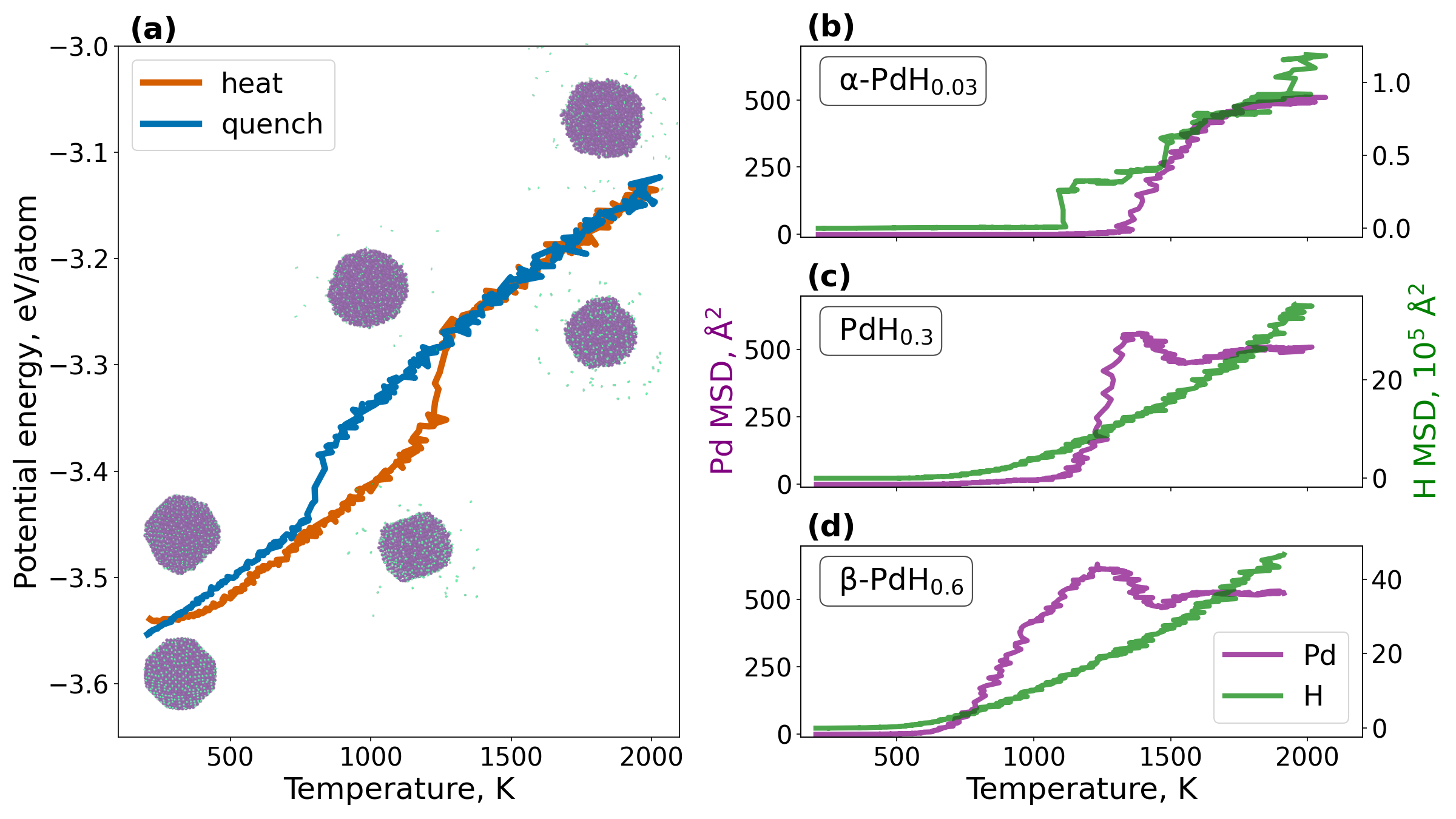}}   
    \caption{\textbf{Rapid melting and quenching of nanoparticles.} (a) Evolution of potential energy per atom during the heating and subsequent quenching cycle for the PdH$_{0.3}$ nanoparticle. Panels (b), (c) and (d) show the mean square displacement (MSD) of Pd and H atoms as a function of temperature during the heating cycle for $\alpha$-PdH$_{0.03}$, PdH$_{0.3}$, and $\beta$-PdH$_{0.6}$ nanoparticles, respectively.
    }%
    \label{fig:nanomelt}%
\end{figure*}

To test the stability of the ACE potential under extreme conditions, we subjected various nanoparticles to rapid melt-quench cycles, heating to 2000 K, and quenching back to 300 K. Each cycle was conducted over 6 nanoseconds, corresponding to an effective heating and cooling rate of approximately 0.6 K/ps. Although such rapid temperature changes exceed physically attainable conditions, these tests provide stringent evaluations of performance under highly non-equilibrium scenarios.

The melt-quench tests were performed on truncated octahedral nanoparticles with varying sizes. For clarity, we only present the results obtained for nanoparticles with an initial diameter of 4 nm.  As in the previous simulation, H atoms were randomly distributed to generate nanoparticles of three distinct H concentrations: $\alpha$-PdH$_{0.03}$, PdH$_{0.3}$, and PdH$_{0.6}$. Our simulations show stable melting of nanoparticles up to 2000 K, followed by recrystallization into roughly spherical shapes primarily bounded by (111) facets upon cooling. 

The evolution of potential energy per atom during the melt-quench thermal cycle for the PdH$_{0.3}$ nanoparticle is shown in Fig.~\ref{fig:nanomelt}(a). During heating, the nanoparticle undergoes desorption of surface-segregated hydrogen as H$_2$ gas molecules, followed by melting under superheated conditions. The transition to a liquid state is marked by a sharp rise in potential energy. During quenching, undercooling leads to a rapid recrystallization of the nanoparticle into a spherical shape, reflected by a sudden energy drop at approximately 900 K. The temperature difference between superheating and undercooling results in a characteristic thermal hysteresis in the potential energy profile.

Figures~\ref{fig:nanomelt}(b)--(d) show the mean square displacement (MSD) of Pd and H atoms at three different concentrations. Melting is indicated by a sharp increase in the MSD of Pd atoms. Notably, the melting temperature of the nanoparticle decreases with increasing H concentration, suggesting that hydrogen content significantly affects the thermal stability of PdH$_x$ nanoparticles.

\section{Conclusions and outlook} \label{sec:conclusions}

Reliable modeling of Pd-H interactions is key for both fundamental surface science and hydrogen storage technology. In this study we introduced a general-purpose ACE potential for the Pd-H system, rigorously validated against a broad range of bulk and surface properties. The ACE potential achieves near-DFT accuracy with an orders-of-magnitude reduction in computational cost, bridging the gap between electronic structure methods and large-scale atomistic simulations.

\begin{figure}[]
    \centering
    {\includegraphics[width=0.97\columnwidth]{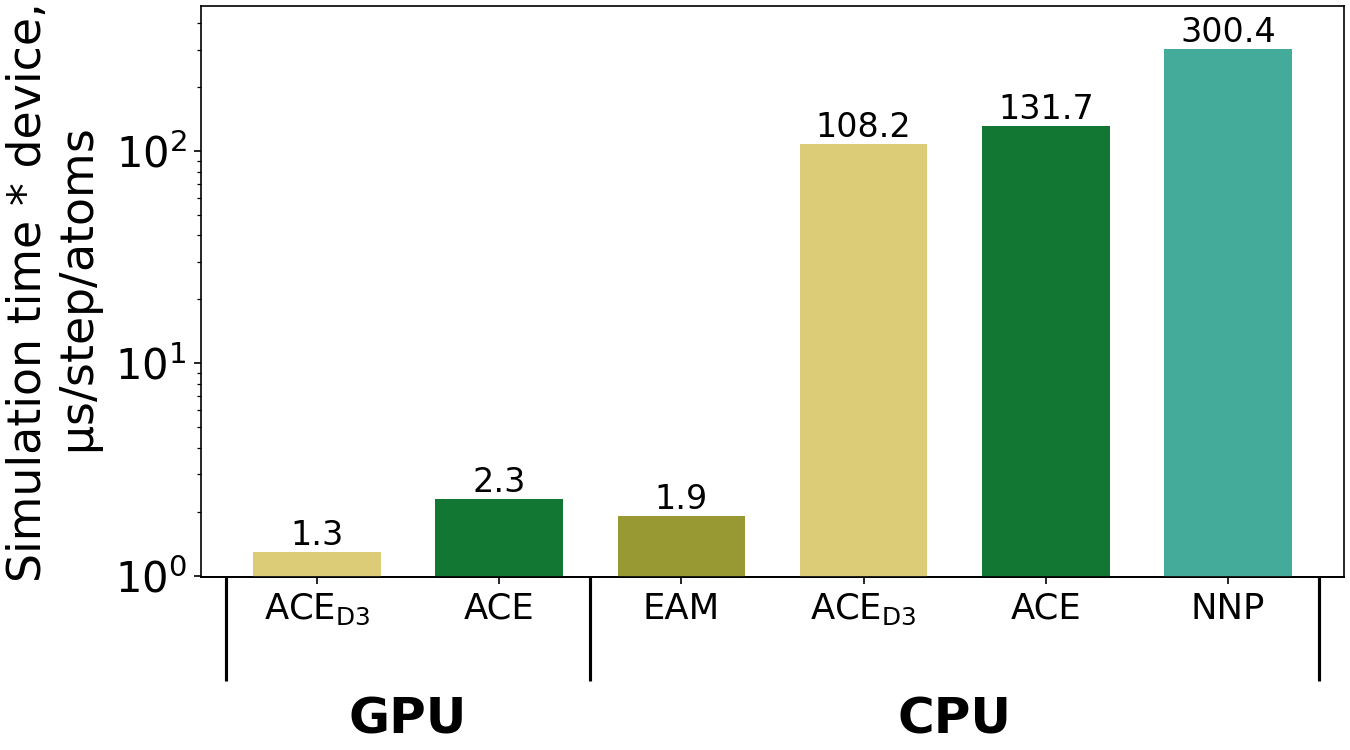}}   
    \caption{Comparison of CPU (AMD Ryzen 5 3600X) and GPU (Tesla V100S-PCIE 32 GB) times for ACE, ACE$_{\text{D3}}$, EAM and NNP models. CPU times are given per core, GPU times are given per device.}%

    \label{fig:comp_timings}%
\end{figure}

Compared to existing interatomic potentials, including the neural network potential by Kimizuka et al.~\cite{kimizuka2022artificial}, ACE consistently matches or exceeds predictive performance across multiple key properties.  Notably, ACE is at least three times faster than the NNP on CPU (see Fig.~\ref{fig:comp_timings}), and achieves microsecond force-calls on a typical GPU, making it exceptionally well-suited for long-time-scale simulations required to observe phase transitions.

We demonstrated the potential's practical applicability through finite-temperature simulations of experimentally relevant PdH$_x$ nanoparticles of varying geometries and sizes. ACE successfully predicts the experimentally observed phase separation into the $\alpha$ and $\beta$ phases, confirming previously proposed core-shell structural models, and accurately predicts lattice parameters across different nanoparticle sizes. Furthermore, the potential's robustness and transferability are highlighted by its stability during rapid melt-quench simulations, underscoring its reliability even under extreme, non-equilibrium conditions.

Overall, this ACE potential represents a significant advancement in the computational modeling of Pd–H systems, offering an optimal balance of accuracy, efficiency, and transferability for future studies of hydrogen-metal interactions.

\section{Methods} \label{sec:moethods}

\subsection{DFT settings for the training data}

The energies and interatomic forces of all the structures in the training dataset are computed using the Vienna ab-initio simulation package (VASP).~\cite{kressabinitio1993,kresse1996efficiency,kresse1996efficient} With the exception of atomic clusters, all structures are computed with a uniform k-mesh density of 0.125 eV $\mathrm{\AA}^{-1}$. A single k-point at the high-symmetry $\mathrm{\Gamma}$ point is used for non-periodic clusters. The cutoff for the number of plane waves is set to 500 eV, a Gaussian smearing with width 0.1 eV is used. The exchange-correlation is approximated using the generalized gradient as given by Perdew, Burke and Ernzerhof (GGA-PBE functional)~\cite{PBEXC}. The convergence limit for the self-consistency loops is set to 10$^{-8}$ for the total energy and an additional support grid for the computation of forces is added (\texttt{.ADDGRID.=True}).

\subsection{Reference training data}\label{sec:training}

\begin{figure*}[]
    \centering
    {\includegraphics[width=0.96\textwidth]{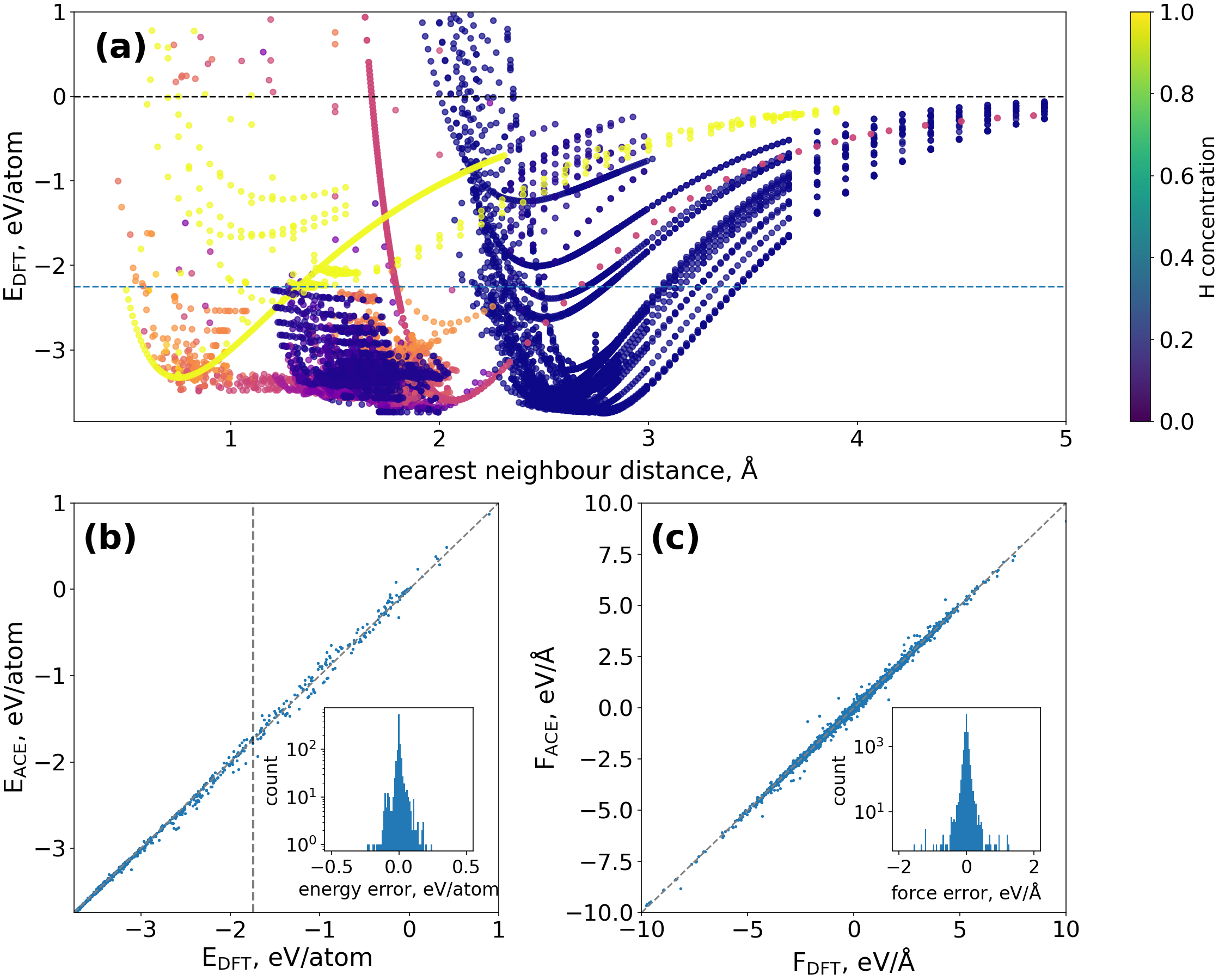}}   
    \caption{ \textbf{DFT reference dataset}
    (a) Complete training dataset for Pd-H system. The points are colored according to the H-concentration. Correlation plots for the 10\% split test set showing (b) predicted energies as a function of reference DFT energies and (c) predicted forces as a function of reference DFT forces on every atom. Insets show the respective error distribution in the test set.
    }%
    \label{fig:refrence_data}%
\end{figure*}

An optimal reference dataset should exhaustively sample the entire PES. For a binary system, this entails sampling of not only a broad range of atomic positions but also compositions. Since the developed potential is intended for studies of bulk Pd-H phases, the main focus was given to Pd and Pd-H systems while only the most relevant H configurations were explicitly included. These included the H$_2$ molecule and a small number of bulk phases, such as fcc, hcp and bcc. Additional H-H interactions were sampled indirectly via Pd-H binary configurations, which often contain multiple interacting H atoms, for instance, at Pd surfaces or in bulk crystal defects. 

A large number of bulk Pd structures were included in the dataset, including standard phases like fcc, hcp, dhcp, bcc, simple-hexagonal, and diamond-cubic, among others. Large volume variations are included with a high density of data near the equilibrium volume and a low density of points at larger volumes. Additionally, various configurations of the ground state fcc phase are thoroughly sampled. These include defect structures such as mono-vacancy, di-vacancy, grain boundaries and surface configurations corresponding to different orientations. Additionally elastically deformed unit cells are added to sample the elastic constants and supercells containing near-equilibrium finite displacements of atomic positions are added to properly sample the phonon spectra.  

For the Pd-H binary system, the long-ranged volume variations of the $\beta$ phase are included with a high density of data-points. In addition, small volume variations are also included for a few other binary phases. The majority of the binary data is focused on both $\alpha$ and $\beta$ hydride phases. A number of Pd-supercells containing varying concentrations of H atoms are considered with extensive variations in their atomic-positions and volumes. Similar snapshots are included for PdH crystals with vacancies as well as surfaces with different orientations. A collection of ab-initio molecular dynamics (AIMD) snapshots is also included for Pd and Pd-H surfaces with varying concentration of H. In addition to the metal hydride structures, important interactions of single H atom (or a small cluster of H atoms) with bulk Pd and its defects are included. These are migration paths of an H atom from octahedral interstitial site to a tetrahedral site in bulk Pd, surface migration of H atom over different trapping sites and migration of a surface H atom to subsurface layers for the important surface orientations, and clusters of up to 7 H atoms in a single vacancy in bulk Pd.

A first version of the potential, parameterized using this initial dataset is then employed to run various small-scale MD simulations. These simulations include PdH$_{x}$ surfaces with varying $x$ and temperature ranges of up to 2000 K, which is higher than the melting temperature of Pd. We then employed active-learning (AL) to select extrapolated and representative structures from these MD-runs. This involves the calculation of the extrapolation grade ($\gamma$) associated to every structure based on the D-optimality criterion~\cite{lysogorskiy2023active}. Structures with values of $\gamma$ larger than 8 were selected for training. Additionally, the ab-initio random structure search (AIRSS)~\cite{Pickard_2011} package was employed to generate pure Pd and PdH random structures. The most representative of these random structures were selected via AL and added to the reference data. 

The computed energies per atom as a function of nearest-neighbour distance for all collected structures are shown in Fig.~\ref{fig:refrence_data}(a). In total, the reference dataset comprises 13,127 structures, including 444 pure H structures and 5,769 pure Pd structures. The remaining 6,914 structures correspond to binary Pd--H configurations.

\subsection{Parameterization}

\begin{table}[]
\caption{ACE configurations for both the fitted potentials. $r_{cut}$ and $r_{in}$ are given in $\AA{}$ and correspond to the outer and inner cutoffs, respectively.}
\label{tab:pot_config}
\begin{tabular}{lccccc}
\hline
Interaction & $r_{cut}$ & $r_{in}$ & $n_{max}$ & $l_{max}$ & \begin{tabular}[c]{@{}c@{}}No. of \\ functions\end{tabular} \\ \hline
\multicolumn{6}{l}{\textbf{ACE}}                                                                                                                   \\
Pd--Pd & 6.5 & 1.2  & 16/5/4/2 & 0/4/2/2 & \multirow{3}{*}{938} \\
Pd--H  & 6.0 & 0.9  & 20/3/2/1 & 0/4/2/2 &                        \\
H--H   & 4.0 & 0.45 & 20/1     & 0/1     &                        \\ \hline
\multicolumn{6}{l}{\textbf{ACE$_{\text{D3}}$}}                                                                                                     \\
Pd--Pd & 6.5 & 0.45 & 11/5/4/2 & 0/4/2/1 & \multirow{3}{*}{385} \\
Pd--H  & 6.0 & 0.45 & 9/3/3/1  & 0/4/2/2 &                        \\
H--H   & 4.0 & 0.45 & 20/3/2/1 & 0/4/2/2 &                        \\ \hline
\end{tabular}
\end{table}

The \texttt{PACEMAKER}~\cite{bochkarevefficientpara} package was used to parameterize an ACE potential over the reference dataset. Simplified Bessel function was used to model the radial basis functions and an energy-based-weighting scheme was applied so that all structures above 2.5 eV of the convex hull are given 95\% higher relative weights. This energy cutoff is marked by the horizontal dotted line in Fig.~\ref{fig:refrence_data}. Additionally, important selected structures were manually given a higher weight.
90\% of the reference data is used for training the model and 10\% is split as the test set. The correlation plot of the split test data for both energies and forces are shown in Fig.~\ref{fig:refrence_data}. Final RMSE of the model are 24.95 meV/atom in energies (MAE: 10.14 meV/atom) and 67 meV/\AA{} in forces (MAE: 30 meV/\AA{}). 

The ACE expansion uses 938 basis functions, providing sufficient accuracy across all relevant properties. The cutoffs are chosen separately for different interactions: Pd--Pd bonds use a cutoff of 6.5 \AA{}, Pd--H bonds 6.0 \AA{}, and H--H bonds 4.0 \AA{}. The inner cutoff ($r_{\mathrm{in}}$) for each interaction defines a short-range distance below which an additional repulsive term is introduced to prevent unphysical atomic overlap and ensure numerical stability.

The maximum number of basis functions $n_{max}$ for the Pd--Pd interaction is set to 16 while that for the H--H interaction is 20. Both Pd--Pd and Pd--H interactions are modelled up to 5 body-orders (see Table~\ref{tab:pot_config}) while the H--H interaction is modelled mainly with a pair-wise term with a third body-order having only 1 basis function. The configuration shown in Table~\ref{tab:pot_config} reflects the complexity of interactions the elemental pair contains. While H can exist in bulk phases under extreme conditions, under the conditions considered for the current model, H exists exclusively as H$_2$ gas.

\subsection{LAMMPS}

All molecular dynamics (MD) simulations in this study were performed using the Large-scale Atomic/Molecular Massively Parallel Simulator (LAMMPS) package \cite{plimpton1995fast, LAMMPS2}. A timestep of 1 fs was employed throughout all simulations. GPU-accelerated simulations were carried out using the KOKKOS package within LAMMPS \cite{trott2022kokkos, CarterEdwards20143202}. The conjugate gradient algorithm is used for ionic optimization, and the thermostat and barostat given by Nosé–Hoover are utilized to control temperature and pressure, respectively \cite{nose1984unified, hoover1985canonical}.

\subsection{Data availability}

The atomic cluster expansion potentials, and the corresponding datasets, generated in this study are openly provided in the Zenodo repository under \url{https://doi.org/10.5281/zenodo.20591788}. These comprise the PBE-based ACE potential (\texttt{PdH\_PBE\_v03a.yaml}, together with its active-set file \texttt{PdH\_PBE\_v03a.asi} for on-the-fly extrapolation-grade evaluation) and the PBE+D3-based ACE$_{\text{D3}}$ potential (\texttt{pdH\_D3.yaml} and \texttt{pdH\_D3.asi}). A self-contained LAMMPS example is also included, comprising an input script and the initial configuration of a $\beta$-PdH$_{0.6}$ nanoparticle (1103 Pd and 662 H atoms) that performs energy minimisation followed by a short NVT equilibration at 300\,K.

\subsection*{Code availability}
The ACE potentials were parameterised using the open-source \texttt{pacemaker} package.~\cite{bochkarevefficientpara} Molecular dynamics simulations were carried out with LAMMPS~\cite{plimpton1995fast, LAMMPS2} using the ML-PACE (\texttt{pace}) pair style. The LAMMPS input script needed to reproduce the example simulation is provided with this article.

\subsection{Acknowledgements}

The authors thank the Irish Centre for High-End Computing (ICHEC), the UK National Supercomputing Service for the provision of computational facilities. M.V. acknowledges funding from the European Union’s Horizon 2020 research and innovation programme (HERMES; 952184), Research Ireland under the Frontiers for the Future Program (MEM-E-TECH; 23/FFP-A/12221), and the Irish Research Council for a Government of Ireland Postgraduate Scholarship to A.N. (MHSVAC; GOIPG/2021/867). A.N. and M.V. thank Prof. Volker Deringer for helpful discussions during the initial stages of the PdH$_{x}$ potential development with Gaussian Approximation Potentials (2021-2022).

M.M. and R.D. acknowledge funding through German Research Foundation (DFG), in the framework of CRC 1625 project A03, project number 506711657.

\bibliography{references} 

\end{document}